\documentclass[review]{elsarticle}

\usepackage{lineno,hyperref}
\usepackage{graphicx}  % needed for figures
\usepackage{dcolumn}   % needed for some tables
\usepackage{bm}        % for math
\usepackage{verbatim}   % for math
\usepackage{textcomp}
\usepackage{amssymb}
\usepackage{amsmath}
\usepackage{epsfig}
\usepackage{tabularx}
\usepackage{xcolor}

\journal{NIM A}

\begin{document}

\begin{frontmatter}

\title{Tracking Algorithms for TPCs using Consensus-Based Robust Estimators}

\author{J. C.~Zamora \corref{cor1}}
\cortext[cor1]{Corresponding author}
\ead{cardona@if.usp.br}
\author[]{G. F.~Fortino}
\address{Instituto de Fisica, Universidade de Sao Paulo, Sao Paulo 05508-090, Brazil}

\begin{abstract}
A tracking algorithm\footnote[1]{Code available at \url{https://github.com/jczamorac/Tracking_RANSAC.git}}  based on  consensus-robust estimators was implemented for the analysis of experiments with time-projection chambers. In this work, few  algorithms beyond RANSAC were successfully tested using experimental data taken with the AT-TPC, ACTAR and TexAT detectors. The present tracking algorithm has a better inlier-outlier detection than the simple sequential RANSAC routine. Modifications in the random sampling and clustering were included to improve the tracking efficiency. Very good results were obtained in all the test cases, in particular for fitting  short tracks in the detection limit. 
\end{abstract}


\end{frontmatter}

% \linenumbers

\section{Introduction} 
Time Projection Chambers (TPC) are among the most efficient  devices existing for charged-particle tracking. They enable measurements of energy loss and position hits in the 3-dimensional space which are employed for particle identification and momentum reconstruction. TPCs have been successfully used during the last three decades as tracking systems in many particle-physics experiments, e.g., TOPAZ  \cite{KAMAE1986423},  STAR \cite{ACKERMANN1999681} or ALICE \cite{ALME2010316}. \par
In the past few  years, the operation of TPCs in Active Target (AT) mode have gained great attention for investigation in  nuclear physics. The AT principle brings the advantage to use a gas as detector material and target medium simultaneously. These detection systems enable complete measurements of nuclear reactions with a large solid angle coverage and low-energy detection thresholds. The target thickness can be increased to achieve measurements at high luminosities without any significant impact in the reconstruction of the  reaction vertex and angular and energy resolutions. Therefore, TPCs are a very powerful technique which is ideal for nuclear reaction experiments involving  low-intensity unstable beams. \par
Currently, many facilities around the world are putting  great effort in the development of active target TPCs as a fundamental part for future research programs \cite{HEFFNER201450,FURUNO2018215,MAUSS2019498, SHANE2015513, BRADT201765, KOSHCHIY2020163398}. In particular, AT-TPC \cite{BRADT201765} (Active Target Time  Projection Chamber), ACTAR-TPC \cite{MAUSS2019498} (ACtive TARget and Time Projection Chamber) and TexAT \cite{KOSHCHIY2020163398} (Texas Active Target)  are devices already in operation. Independent of the geometry or number of channels, the principle of these detector systems is similar. The beam particles impinge the TPC active volume and induce nuclear reactions along their path. The reaction products ionize the gas atoms while traversing the active volume and generate  electrons. Upon applying an uniform electric field, ionization electrons produced by charged particles along their tracks drift towards a MICROMEGAS (Micro-MEsh GAseous Structures) \cite{GIOMATARIS199629} sensor plane at a constant velocity.  Usually, the MICROMEGAS detector is highly segmented ($\sim 10^4$ pads) and allows a precise determination of the energy loss and a reconstruction of a 2D image  of the particle trajectory, while the third dimension  is extracted from the drift time of the electrons. This enables to record a collection of position-energy hits in the space for each event. In order to analyze the 3D images and extract the respective  information of the nuclear reaction, sophisticated tracking algorithms are employed. Depending on the AT properties and the application of the detector setup, the tracking algorithm can employ different pattern recognition methods.  For instance,  Hough Transform \cite{BRADT201765}, Hyperbolic SecantSquared \cite{ROGER2011134}, Kalman Filter \cite{LEE2020163840} and Hierarchical Clustering \cite{DALITZ2019159} are  few  methods that are used for tracking in TPCs. A comprehensive review of the technology and tracking algorithms used for TPCs can be found in Ref.~\cite{yassidrev}. \par

Recently, a tracking algorithm based on RANSAC  (RANdom SAmple Consensus) has been developed to analyze data from the AT-TPC \cite{AYYAD2018166}, with a great success in different works \cite{PhysRevLett.123.082501}. RANSAC \cite{10.1145/358669.358692} is a very popular robust estimator algorithm in computer vision with a wide range of applications, e.g., multi-model fitting \cite{6873258}, image mosaicing \cite{10.1016/j.jvcir.2015.10.014}, structure of motion \cite{1238341} and many others \cite{hartley2003multiple}. However, RANSAC is rather sensitive to the inlier-outlier threshold and it requires a prior parameter fine tuning  each time for a specific application \cite{1640543}. Also, a common problem of the algorithm is to fail when describing data that contain multiple structures, and the reason is because RANSAC is designed to extract a single model \cite{608280}.  Several variants of RANSAC have been proposed to improve the performance of the algorithm \cite{1640543,10.1007/978-3-540-88688-4_37}.  Most of these strategies aim to optimize the process of model verification by using probabilistic approach or modifying the sampling process in order to preferentially generate more useful hypotheses. For example, MLESAC (Maximum Likelihood Estimation SAmple Consensus) \cite{TORR2000138}, LMedS (Least Median of Squares) \cite{doi:10.1080/01621459.1984.10477105} and J-Linkage \cite{10.1007/978-3-540-88682-2_41} are three common robust estimators employed when RANSAC is not sufficient.\par
In this work, a comparative analysis of tracking algorithms based on the above mentioned robust estimators is performed. To test the performance of the codes, experimental data from the active targets AT-TPC, ACTAR and TexAT  are used. The present article is organized as follows: in Section 2, we give a brief overview of the different algorithms implemented in this work, Section 3 shows their performance with the test data sets, and in Section 4 are presented the conclusions.

\section{Robust Estimators}
\subsection{RANSAC}
RANSAC is an algorithm for robust fitting of models in the presence of many data outliers \cite{10.1145/358669.358692}. The principle of the algorithm is based in two parts:  generating a hypothesis from random samples and verifying it to the data. The hypotheses generation consist in a random selection of a subset of $n$ points to generate a model. In the case of a 3D line ($n=2$),   the model can be parameterized as
\begin{eqnarray}
x&=&x_0 +ta, \nonumber \\
\label{3dli}
y&=&y_0 +tb,\\
z&=&z_0 +tc, \nonumber
\end{eqnarray}
for $t \in (-\infty,\infty)$, where $(x_0,y_0,z_0)$ is a point on the line and $\overrightarrow{v}=\langle a,b,c\rangle$ is a vector parallel to the line.  For  TPCs using a constant magnetic field such as AT-TPC \cite{BRADT201765} or  SPECMAT \cite{specmat}, the trajectory for charged particles can be approximated to a helix function
\begin{eqnarray}
x&=&r\cos(t), \nonumber \\
y&=&r\sin(t),\\
z&=&ct, \nonumber
\end{eqnarray}
for $t \in (0,2\pi)$, where $r$ is the helix radius and $2\pi c$ is the separation between helix loops. In this case, the number of points required for the random sampling is $n=3$, but the model might also need to constraint the radius with a cylinder \cite{helix}.\par
In the second step of RANSAC, each of the hypotheses generated is tested interactively in order to maximize the inliers ratio ($n_\text{inlier}/N_\text{tot}$) by using the cost function
\begin{equation}
 C_\text{RANSAC}(\varepsilon^2)=
\begin{cases}
0 & \varepsilon^2<T^2\\
\text{const.} & \varepsilon^2\geq T^2,\\
\end{cases}
\end{equation}
where $T$ is a threshold allowing to judge whether a given point is an inlier and $\varepsilon$ is the minimum distance (error) between the model and a point. Thus, the inliers scores nothing and each outlier scores a constant penalty. However, RANSAC needs to adjust the threshold parameter which is rather sensitive to accuracy. If $T$ is sufficiently large, then all the matches would be inliers, but if $T$ is too small the algorithm finds several solutions. After optimizing the parameters, the resulting best model from RANSAC can be improved by fitting the inliers with the least squares method \cite{AYYAD2018166}. In the case of a linear fit in 3D, the method has an analytic solution \cite{3dfit}.

\subsection{MLESAC}
MLESAC \cite{TORR2000138} is a robust estimator based on  RANSAC that uses the same idea of generating hypotheses from random sampling, but it differs in the way that the models are tested. The algorithm is improved by using a new cost function 
\begin{equation}
 C_\text{MLESAC}(\varepsilon^2)=
\begin{cases}
\varepsilon^2 & \varepsilon^2<T^2\\
T^2 & \varepsilon^2\geq T^2.
\end{cases}
\end{equation}
In this method, the hypothesis is evaluated on basis of an error probability distribution for both inliers and outliers. It models inlier error as unbiased Gaussian distribution and outlier error as uniform distribution as \cite{BMVC.23.81}
\begin{equation}
\label{mlesacp}
 P(\varepsilon) = \gamma \frac{1}{\sqrt{2\pi \sigma^2}} \exp{\left(-\frac{\varepsilon^2}{2\sigma^2} \right)} +(1-\gamma)\frac{1}{\nu},
\end{equation}
where $\gamma$ is the prior probability to be an inlier (related to the inlier ratio), $\nu$ is the size  of the available error space and $\sigma$ is the standard deviation of the Gaussian noise that usually is set to be $T=1.96 \sigma$. Finally, the best model is obtained from the maximum likelihood of the given data. A C++ version of MLESAC from the RTL robust regression tool was implemented in our tracking algorithm \cite{BMVC.23.81}.

\subsection{LMedS}
The LMedS \cite{doi:10.1080/01621459.1984.10477105} estimator yields the smallest value of the median of the squared errors for the entire data set
\begin{equation}
 \phi = \min \left \{ \text{med}_i(\varepsilon_i^2) \right \}.
 \label{lmeds}
\end{equation}
As there is no analytic solution for the previous equation, the same random sampling routine of RANSAC can be used to generate the hypotheses. In order to give a reliable estimate, the sample set must contain at least 50\%  of inliers \cite{186637}, which is completely fine for a hit pattern in a TPC. Thus, the best model is  extracted from Eq.~(\ref{lmeds}) and the result is improved by fitting the inliers with the least squares method. The LMedS routine used in this work was adapted from the ROBEST library \cite{robest}.

\subsection{J-Linkage}
J-Linkage \cite{10.1007/978-3-540-88682-2_41} is an algorithm optimized to fit data with multiple structures (e.g., several tracks).  The random sampling is constructed in a way that neighboring points are selected with higher probability. If a point $x_i$ has already been selected, then the probability to get a point $x_j$ is
\begin{equation}
\label{nusampling}
 P(x_j|x_i)=
\begin{cases}
\frac{1}{Z}\exp{\left( - \frac{\Arrowvert x_i-x_j\Arrowvert^2}{\sigma^2}\right)} & x_i\neq x_j\\
0 & x_i= x_j,
\end{cases}
\end{equation}
where $Z$ is a normalization constant and $\sigma$ is a scale parameter that is chosen heuristically. Then, a threshold parameter $T$ is used to select the inliers (similar to RANSAC), but now assuming a certain probability for each point relative to a model. The second part of the algorithm consist in an agglomerative clustering to link multiple hypotheses that have a similar result. This is done by measuring the degree of overlap of two elements, $A$ and $B$ (points or clusters), with the Jaccard distance
\begin{equation}
 d_J(A,B) = \frac{\arrowvert A\cup B \arrowvert - \arrowvert A\cap B \arrowvert}{\arrowvert A\cup B \arrowvert}.
\end{equation}
The algorithm proceeds by linking elements with distance smaller than 1 and stops as soon as there are no such elements left. The final model parameters for each cluster of points is estimated by least squares fitting. Our code is based on a C++ version of J-Linkage from Ref.~\cite{Feng10semi-automatic3d}.

\section{ Evaluation of the Algorithms with Experimental Data}

In order to test the performance of the algorithms, experimental data taken with the AT-TPC, ACTAR and TexAT were analyzed. The AT-TPC data are from an experiment with a $\text{H}_2$  beam at 1.5~MeV/u impinging in the AT filled with a ${}^{4}\text{He}$ gas at 600~Torr pressure. The AT-TPC was operated without magnetic field for this commissioning run.
Scattering events were measured with a good precision using the 10240 pad configuration and the electrons drift time along 1~m distance without magnetic field \cite{BRADT201765}. The ACTAR data are from a $\alpha$-source calibration. The gas chosen for this measurement was isobutane at 30~Torr, so that the $\alpha$ particles propagate over the entire active volume and stop in the silicon detectors wall placed at  the end of the TPC. The ACTAR pad plane contains 16384 squared pixels (2~mm side) on an active area of $256\text{~mm}\times 256\text{~mm}$ \cite{MAUSS2019498}.  The TexAT data were taken from an experiment with a ${}^{8}$B beam at 3.2~MeV/u interacting with a P5 [${}^{40}$Ar (95\%) + CH$_4$ (5\%)] gas target at 150~Torr. The geometry configuration of TexAT is similar to ACTAR, but with a lower segmentation on the MICROMEGAS pad plane (1024 read out channels) \cite{KOSHCHIY2020163398}.

\subsection{Reaction vertex and angle reconstruction}
\begin{figure*}[!ht]
\centering
\includegraphics[width=0.45\textwidth]{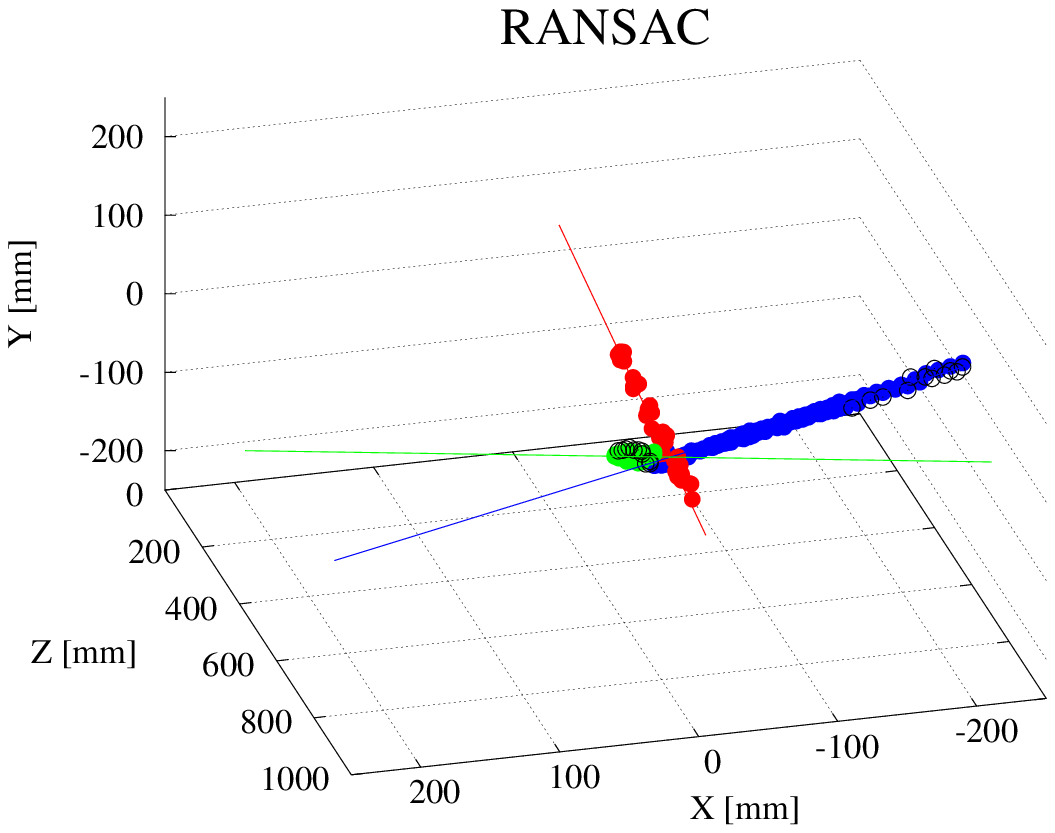}
\includegraphics[width=0.45\textwidth]{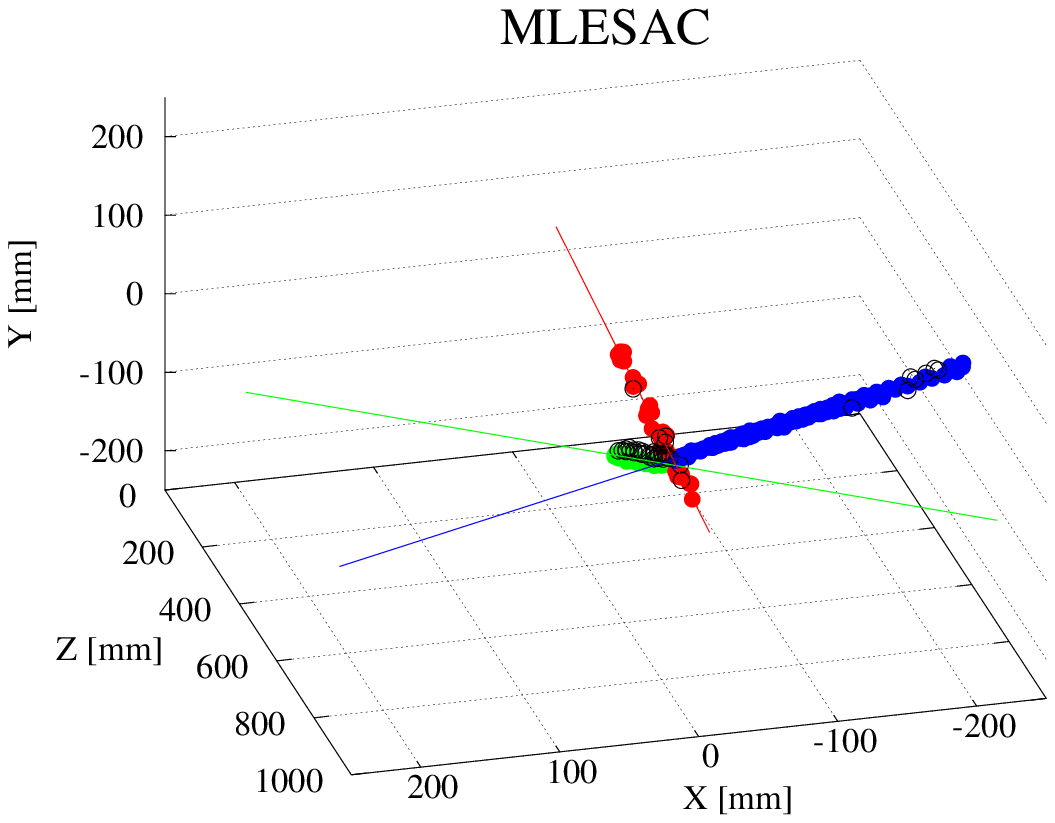}
\includegraphics[width=0.45\textwidth]{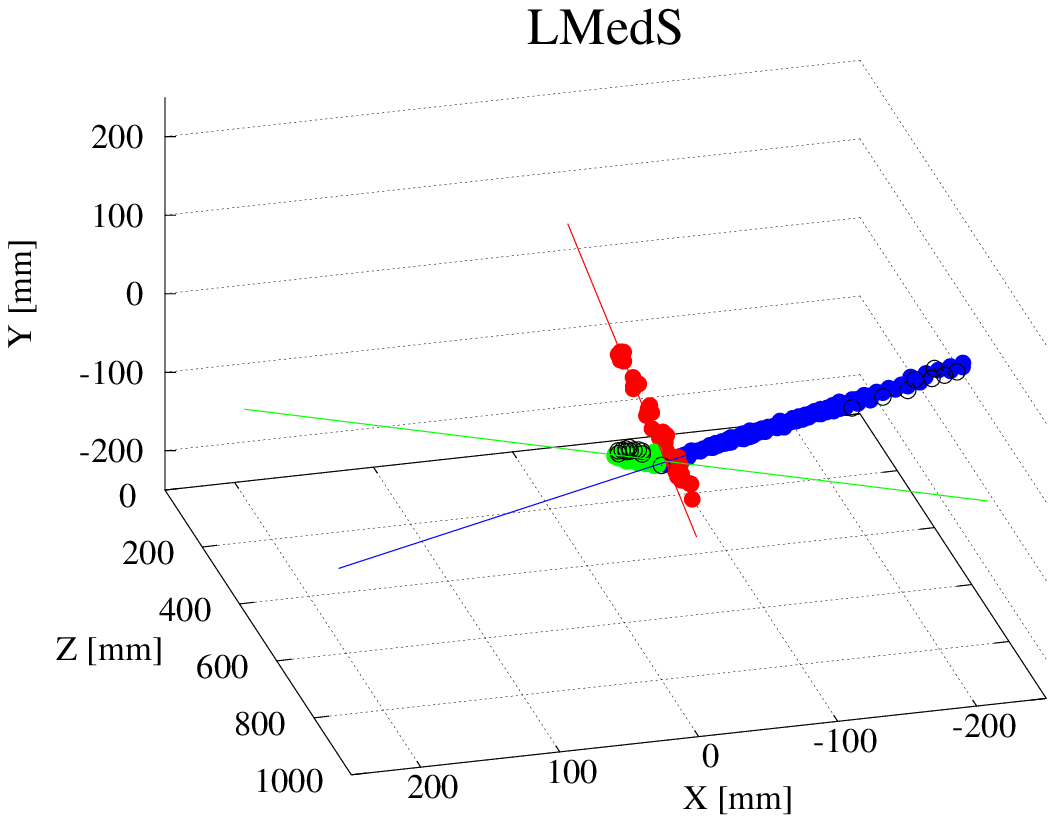}
\includegraphics[width=0.45\textwidth]{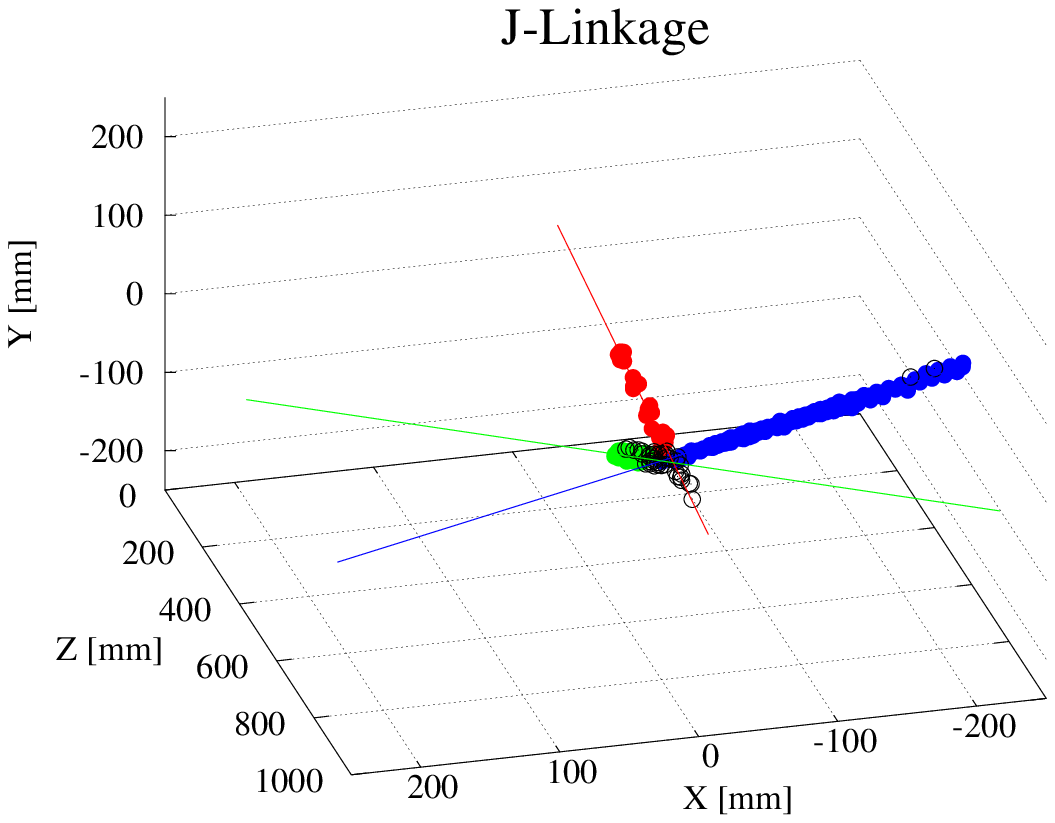}
\caption{\label{attpchits} (color online) Evaluation of the tracking algorithms using experimental data from the AT-TPC (without magnetic field). The colored full points represent a track detected by the algorithm, while the empty points are the outliers (rejected). The lines are a 3D linear fit of the data points for each track. }
\end{figure*}

Scattering events are reconstructed from the 3D hit pattern in the active volume of the TPCs. Each event is composed by one or more 3D linear tracks  (see Eq.~(\ref{3dli})). The intercept point of the beam track with the  other particle tracks corresponds to the reaction vertex.  The objective of the tracking algorithm is to identify and separate the different particle hits within a point cloud and fit them with a certain model. Thus, the kinematics of the nuclear reaction  can be reconstructed from the fitted tracks and the position of the vertex. \\
Consensus robust estimators are very useful for detecting the inlier hits and separate the outliers (e.g., noise or hits of other tracks). The first step of our tracking algorithms  is to identify the different sets of inliers using a random sampling routine to test the models generated with a given robust estimator. This procedure is applied recurrently until the 90\% of the hits of the point cloud were selected or there are no more points left to form a track. The second step corresponds to the minimization of the track model and reconstruction of the reaction vertex. Fig.~\ref{attpchits} shows an example of the analysis for one event of the AT-TPC data using different robust estimators. Each color represents distinct tracks detected by the algorithm, while the empty points are the respective outliers. Although the reconstruction looks very good for all cases, the main difference is the inlier ratio as will be explained later. Given that the model minimization also depends on the inliers and their charge deposited,  small differences in the vertex reconstruction are expected.

\begin{figure*}[!ht]
\centering
\includegraphics[width=0.45\textwidth]{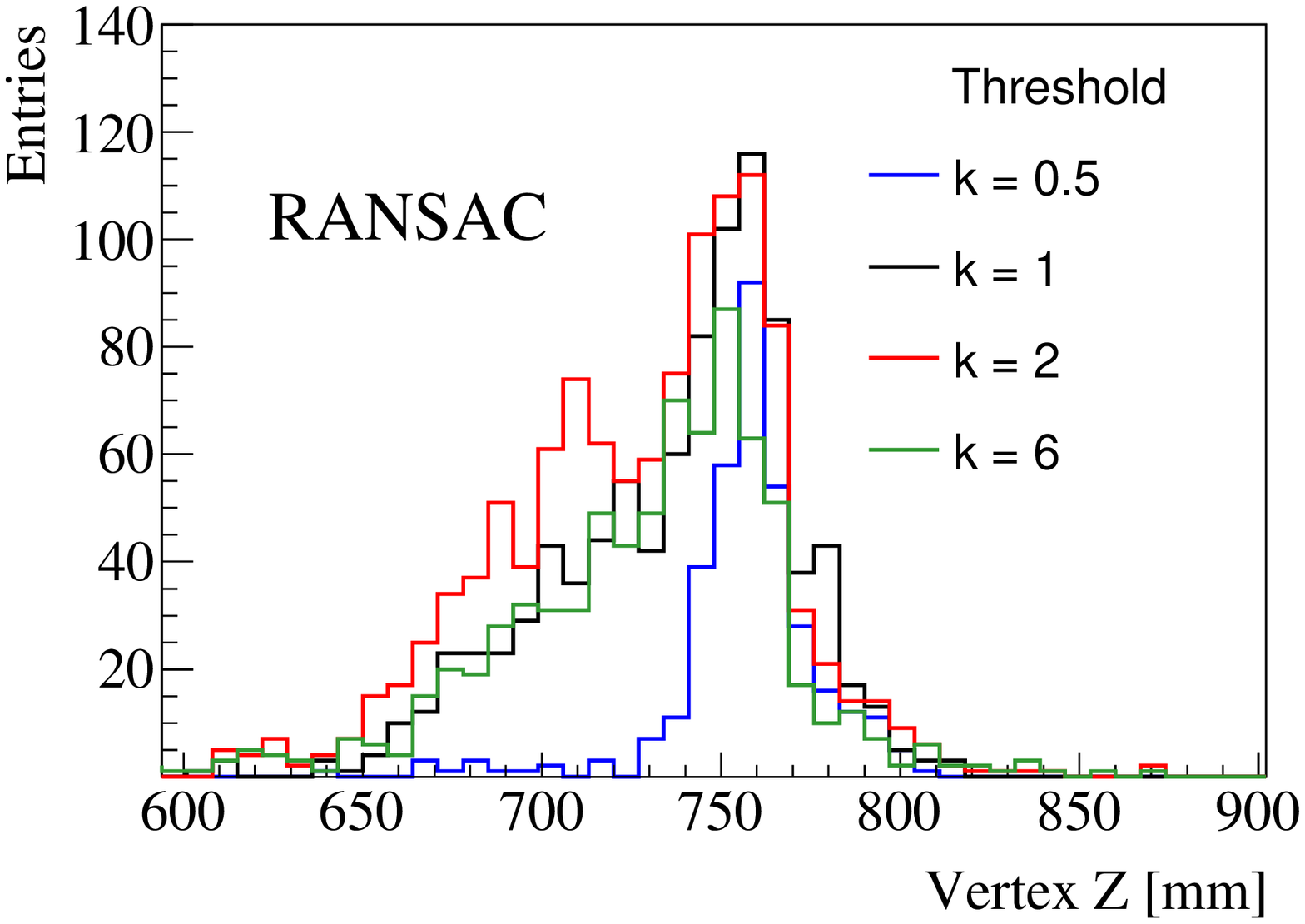}
\includegraphics[width=0.45\textwidth]{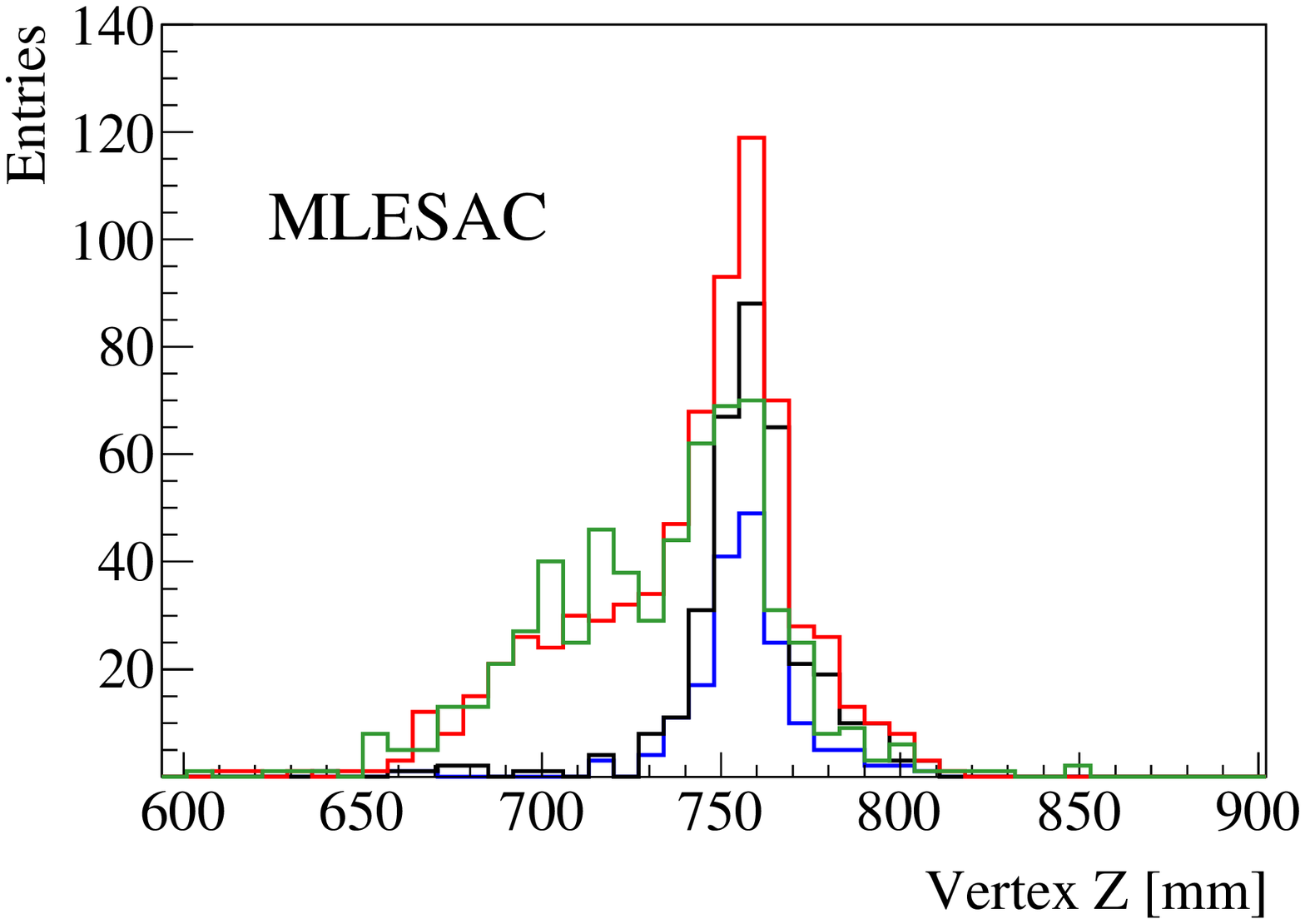}
\includegraphics[width=0.45\textwidth]{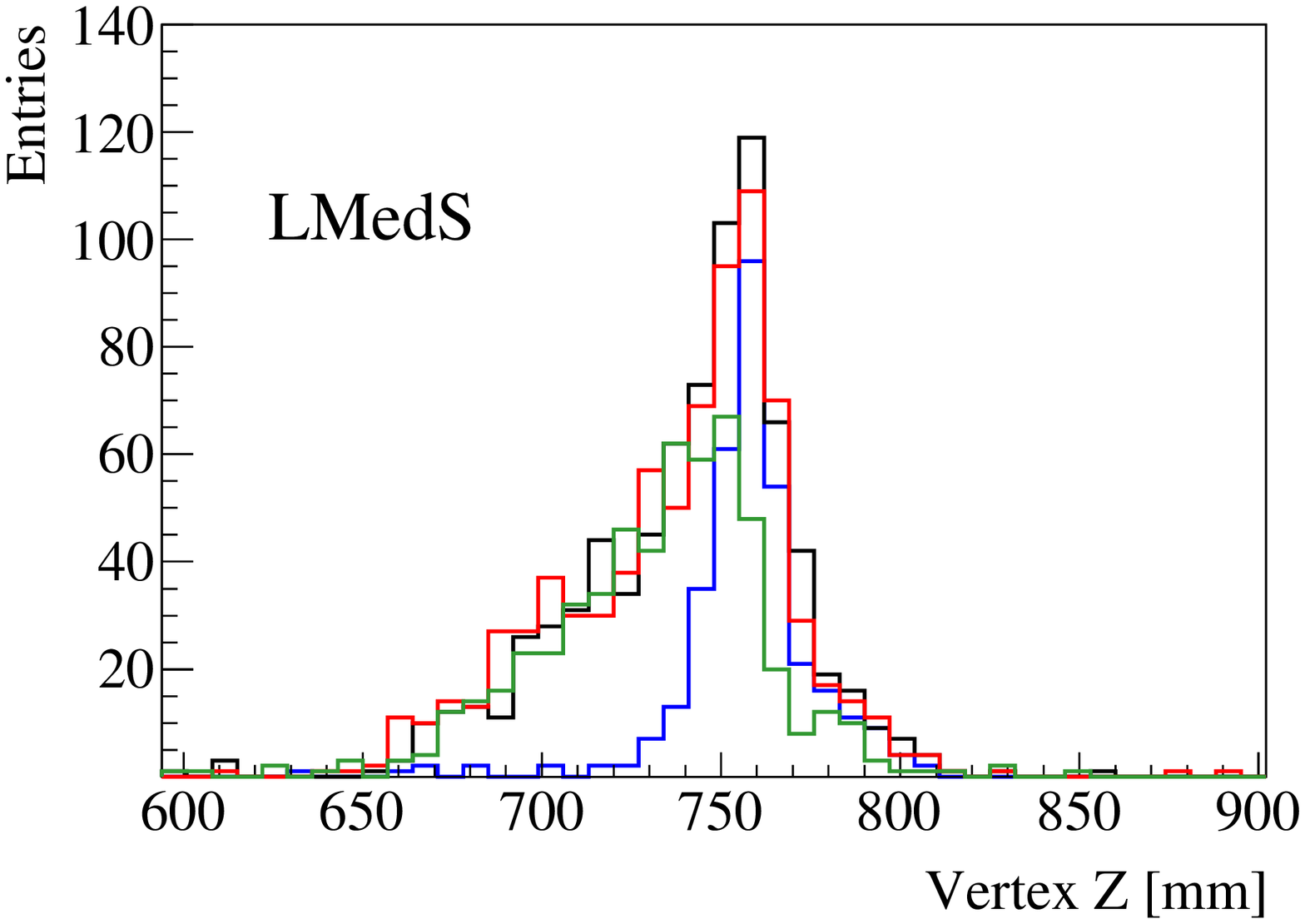}
\includegraphics[width=0.45\textwidth]{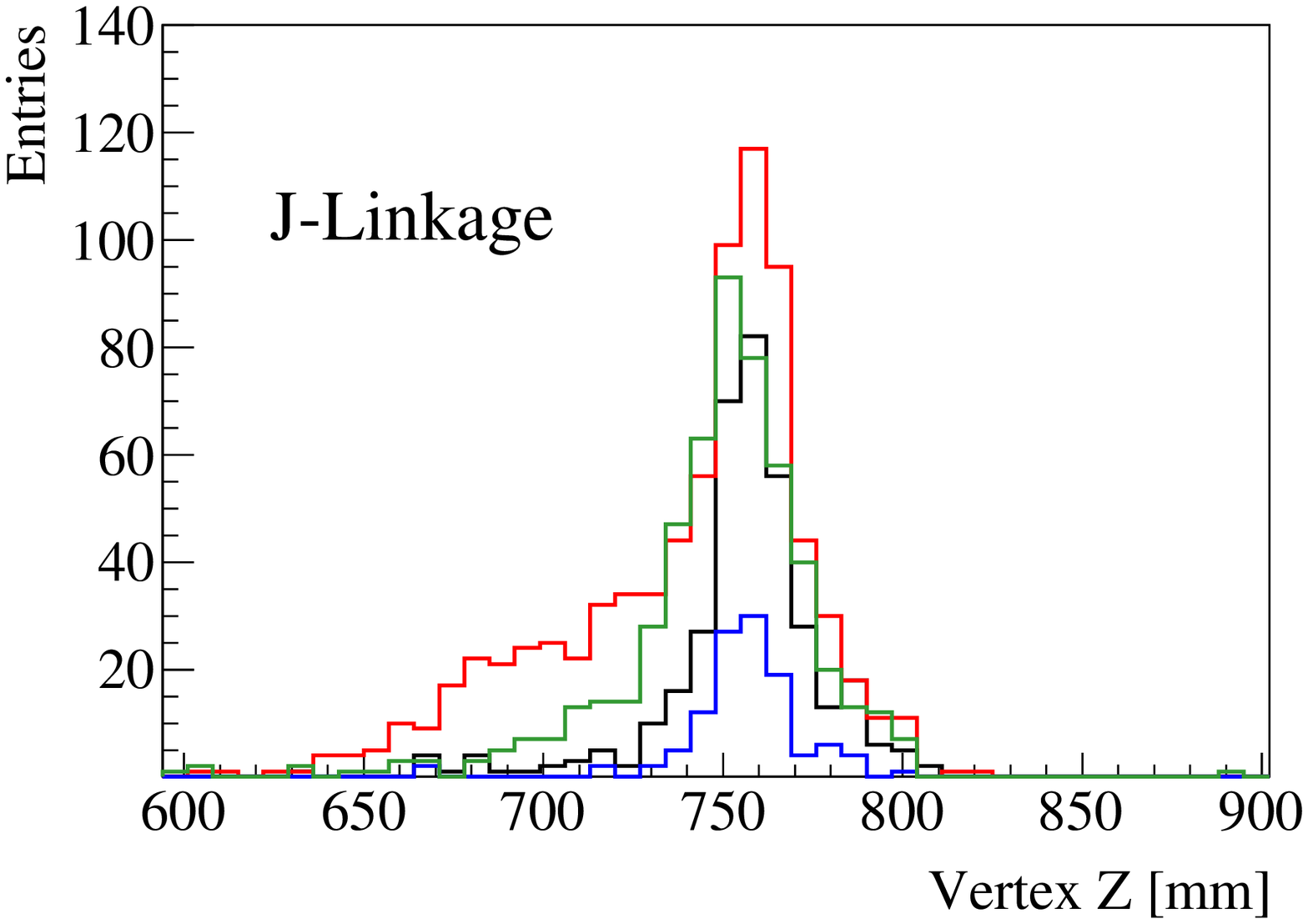}
\caption{\label{vertex} (color online) Projection on the $Z$ axis of the reconstructed reaction vertex. The reconstruction was performed for different distance threshold values: $T=kd$, where $d$ is the pad size.  }
\end{figure*}

\begin{figure*}[!ht]
\centering
\includegraphics[width=0.45\textwidth]{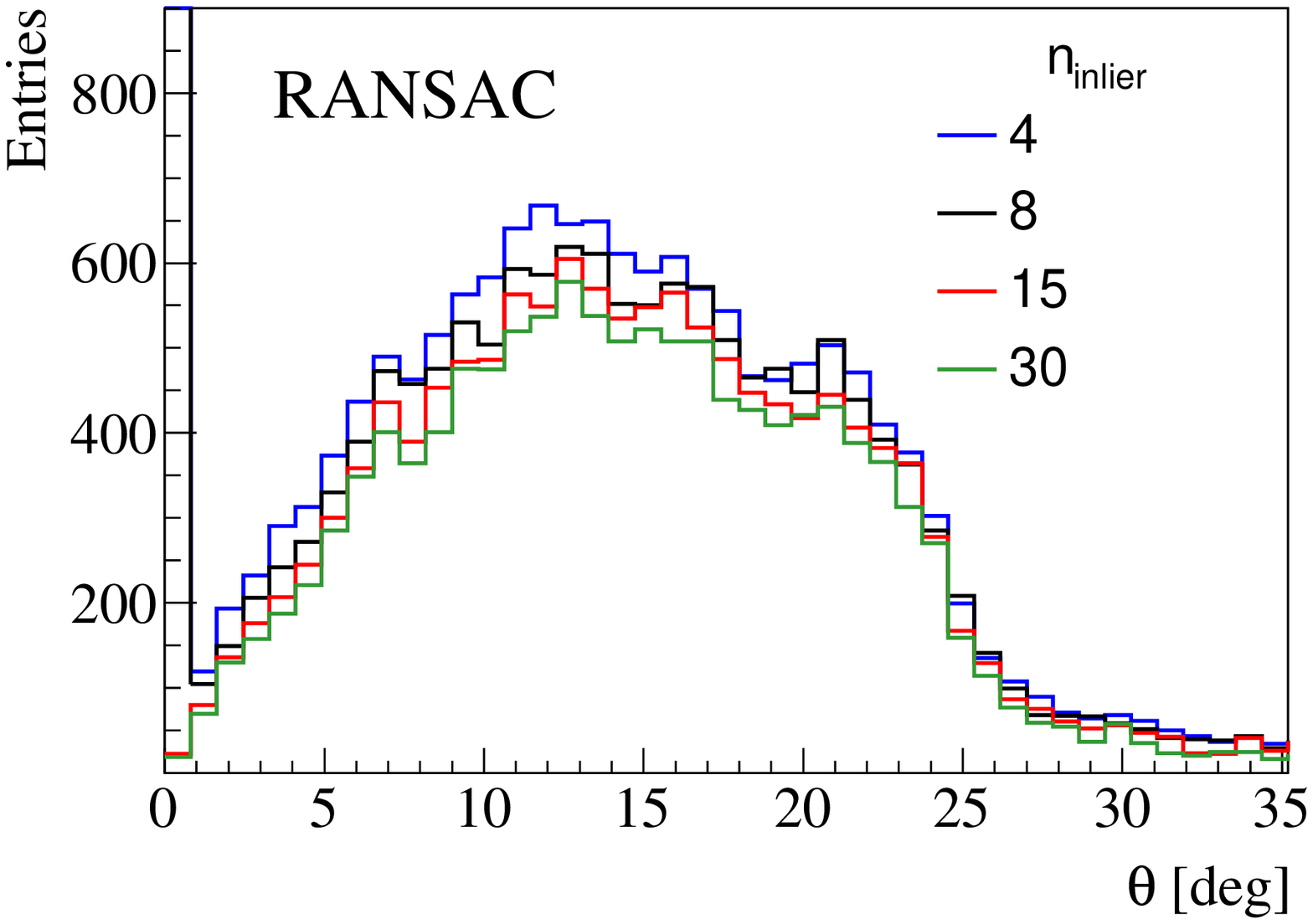}
\includegraphics[width=0.45\textwidth]{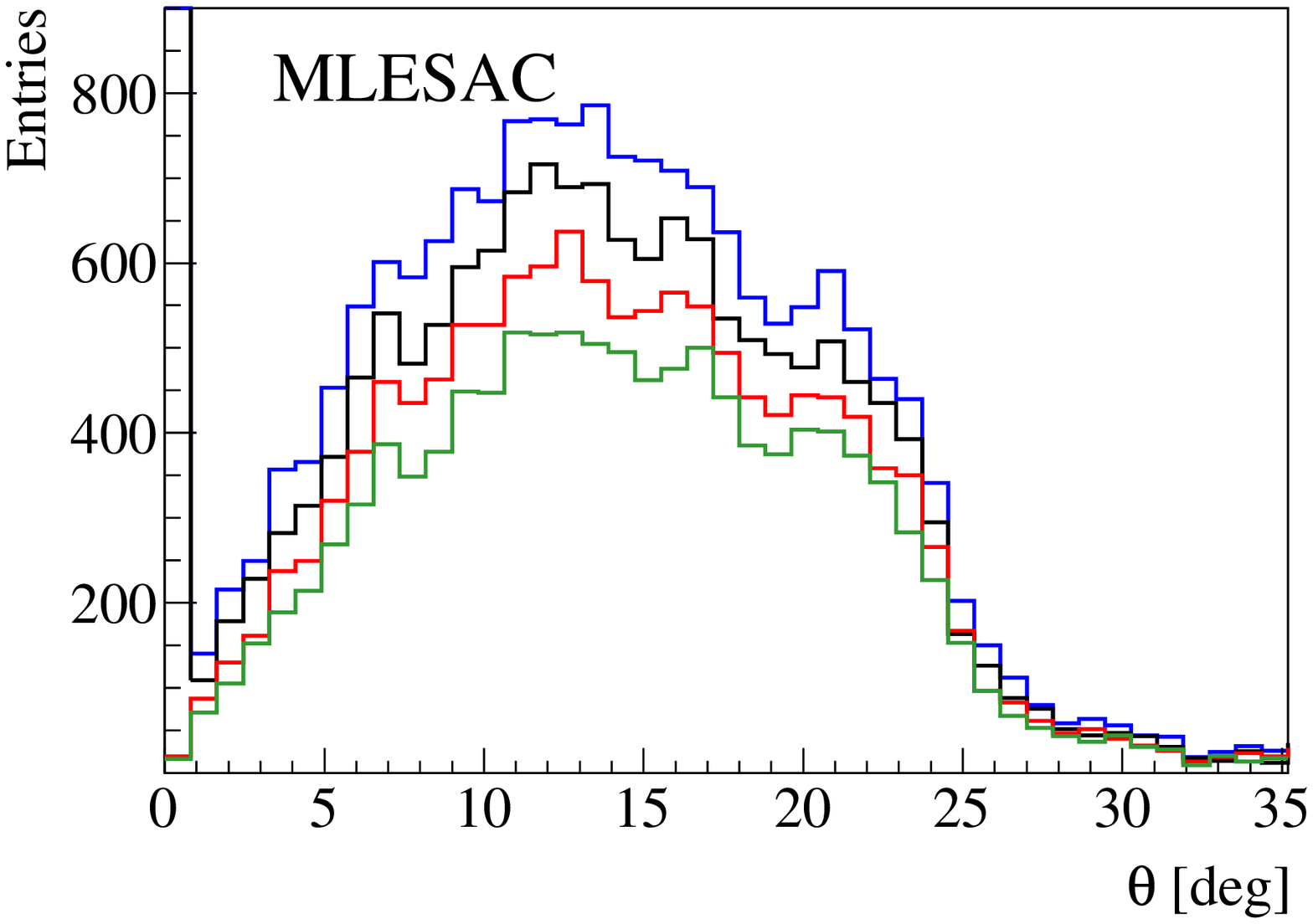}
\includegraphics[width=0.45\textwidth]{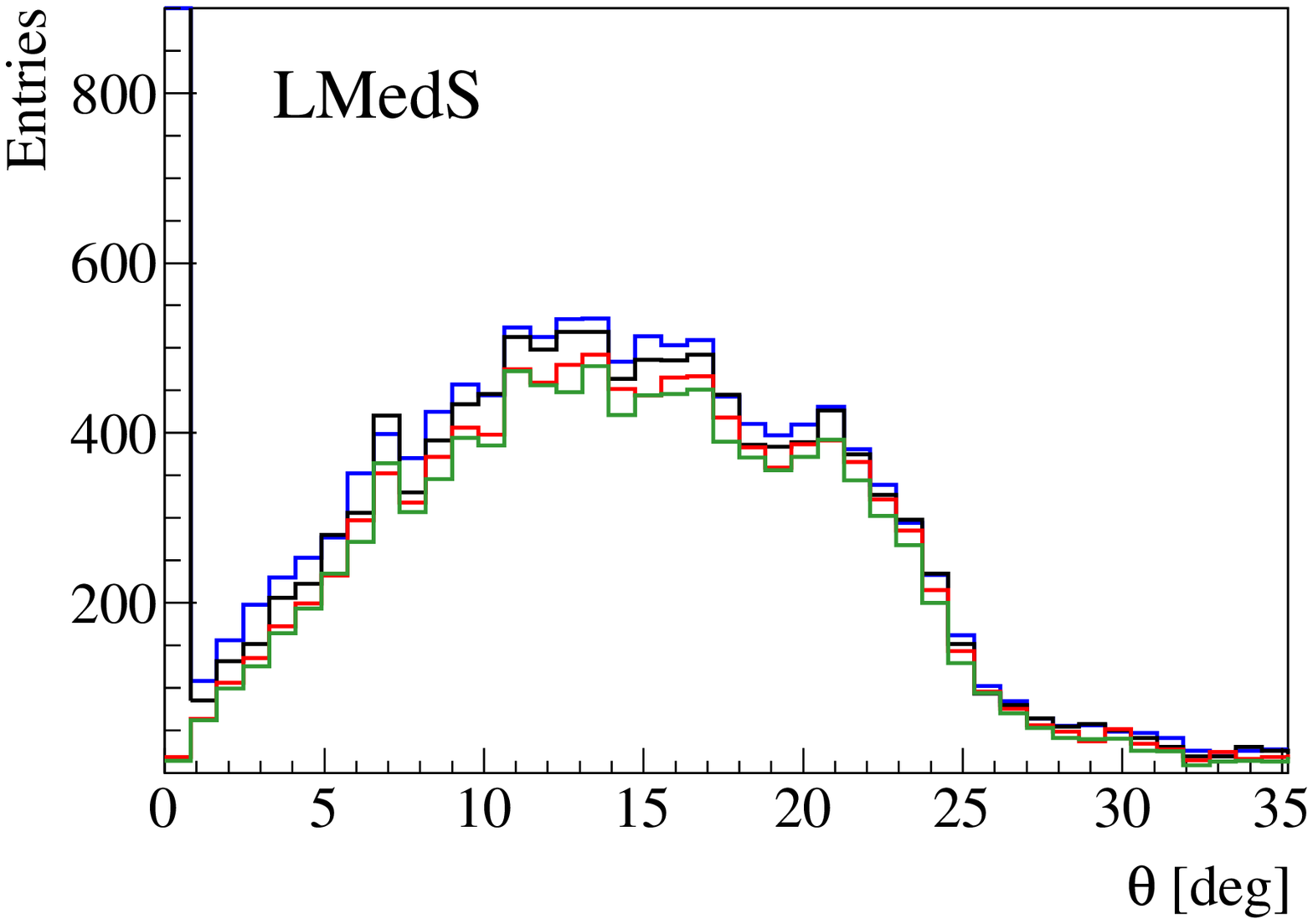}
\includegraphics[width=0.45\textwidth]{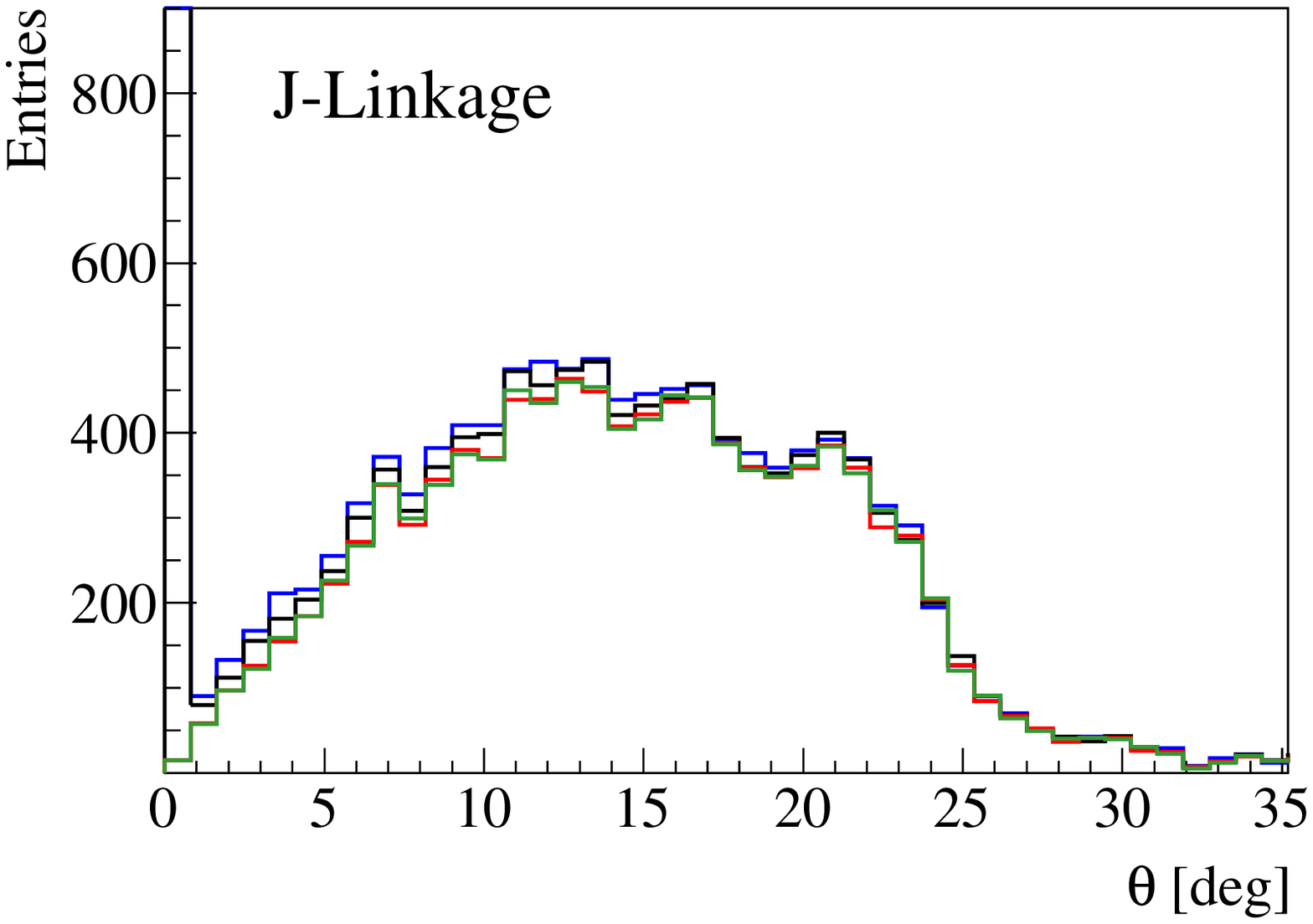}
\caption{\label{actar_ang} (color online) Angle of the tracks relative to the beam axis. The data corresponds to an $\alpha$-source calibration with ACTAR. The reconstructed angle was obtained for a different minimum number of inlier points in each case. }
\end{figure*}

In order to evaluate the vertex reconstruction,  analysis  for several values of the distance threshold ($T$) was performed. Fig.~\ref{vertex} shows the vertex $Z$ coordinate obtained for the four methods. As  each TPC has a different pad size ($d$), the threshold parameter was reduced to $T=kd$, where $k$ had values of 0.5, 1, 1.5, 2, 3, 4 and 6.
The tracking efficiency clearly depends on the threshold used to detect the inliers. Small threshold values (relative to the pad size) lead to a reduced tracking efficiency in all cases. The vertex reconstruction also fails for thresholds above 4 times the pad size, because the algorithms accept as inliers many uncorrelated points (in particular around the vertex region). The best tracking efficiency achieved for RANSAC, MLESAC and J-Linkage was for a threshold of 2 times the pad size, while $k=1.5$ for LMedS. RANSAC seems to be the most sensitive algorithm to the threshold parameter. This is because of the sharp outlier filter that RANSAC uses to test the hypotheses. LMedS is less sensitive to the threshold in comparison to the other methods. As LMedS is an estimator  based on the median value, extra outlier points do not have a strong influence in the result. Another aspect that was evaluated in the analysis was the reconstruction of short tracks, e.g., recoiling heavy particles. Usually, the algorithms fail for  short track detection when the threshold parameter is large. This is because most of the points around vertex region are selected by only one model. Also, small threshold values lead to an wrong multiple structure detection for a single track. The best results for the reconstruction of short tracks were obtained from J-Linkage. As  the random sampling in this algorithm gives a higher probability to neighboring points, small structures from short tracks are efficiently detected. The reconstruction of low-range tracks is particularly important for experiments at  low-momentum transfer in inverse kinematics, where the recoiling-particle tracks are short and with a low density of points. 
\\
The track angle was also extracted in our analysis. The ACTAR data  were chosen to test the angle reconstruction because the $\alpha$ particles from these data have a narrow angular range from 0 to 30~deg. Due to the high granularity of the ACTAR pad plane, it was possible to evaluate the sensitivity with the minimum number of inlier points ($n_\text{inlier}$) for each estimator.
Fig.~\ref{actar_ang} shows the resulting angle reconstruction. As can be noticed, MLESAC is the most sensitive algorithm to the minimum number of inlier points. Different than the other estimators, MLESAC relays on the inlier ratio to generate a guess for the prior probability using the Expectation Maximization method \cite{TORR2000138}. It means that the algorithm requires a $n_\text{inlier}$ large enough to ensure the convergence of the maximum likelihood. The best results were obtained for $n_\text{inlier}>20$. The large yield for the small $n_\text{inlier}$ histograms is indeed an over prediction of the number of models that generate multiple structures in the same track. 
Differently, J-Linkage is almost not sensitive to the  minimum number of inlier points. As the random sampling is optimized to generate more useful hypotheses than the other cases, the particle tracks are efficiently detected regarding the value of $n_\text{inlier}$.\\
Given that the present algorithms use the same linear regression routine for model minimization, the angular resolution is in average the same for all cases. The angle reconstruction depends mostly in the granularity of the detector as well as the length of the tracks. The angular resolution was investigated using simulated data for different particle ranges, as it is presented in Table~\ref{table1}. The range is given in units of the pad size, while the resolution corresponds to the FWHM (full width at half maximum) value. The angular resolution for the shortest tracks (about 11 points) is in the order of $2^\circ$, but for longer tracks the resolution is improved with values down to  $0.3^\circ$. 

\begin{table}[!ht]
\caption{\label{table1} Average angular resolution for different track lengths. The range is given in units of the pad size.   }
\def\arraystretch{1.0}%  1 is the default, change whatever you need
\centering
\begin{tabular}{cc}
\hline \hline 
Range [pad size] &$\Delta \theta_\text{FWHM}$ [deg.]  
      \\ \hline \hline
11(1) & 2.09(4) \\
20(1) & 0.93(2) \\
28(1) & 0.53(1) \\
37(1) & 0.40(1) \\
46(1) & 0.29(1) \\
56(1) & 0.26(1) \\
    \hline \hline    
\end{tabular}
\end{table}

\subsection{Multiple structures}

\begin{figure*}[!ht]
\centering
\includegraphics[width=0.45\textwidth]{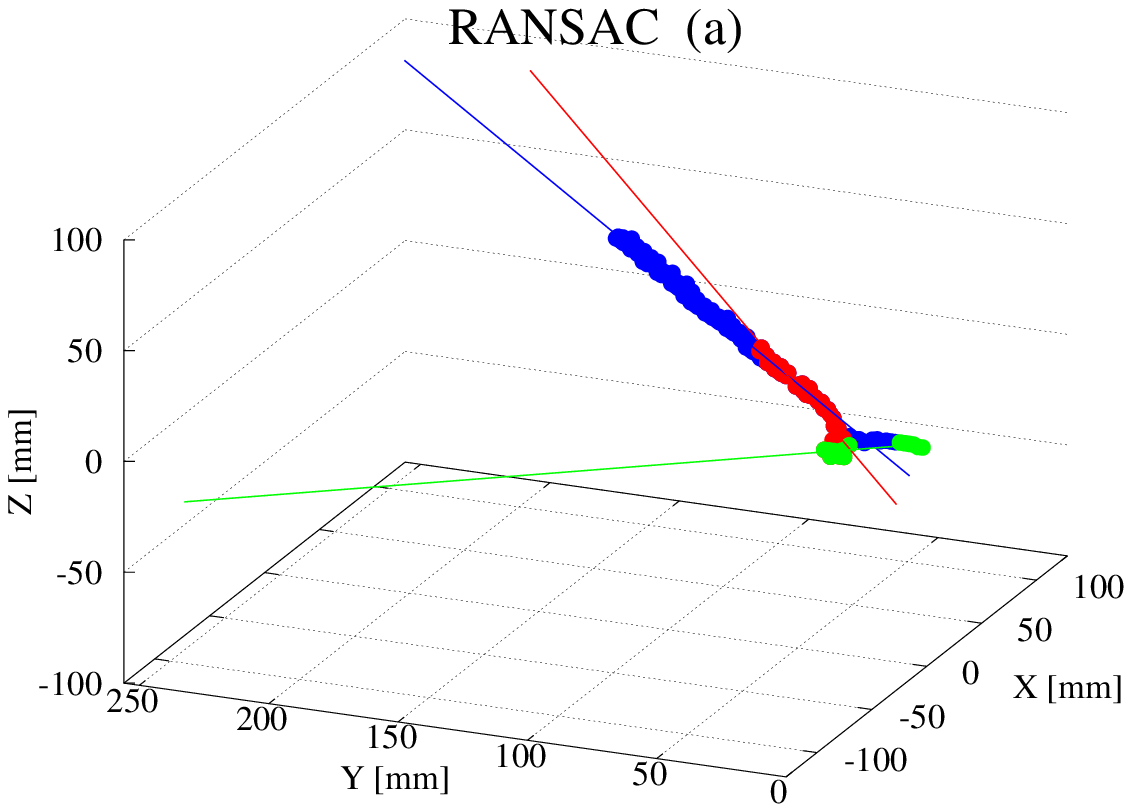}
\includegraphics[width=0.45\textwidth]{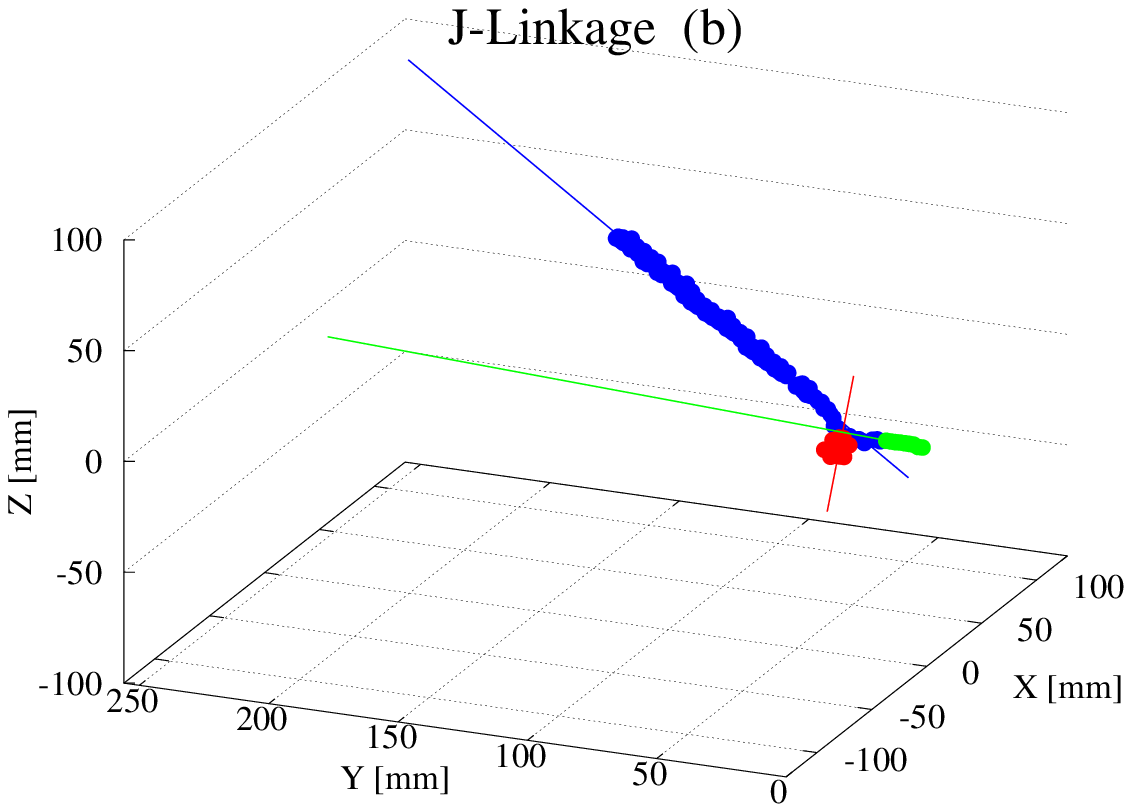}
\includegraphics[width=0.45\textwidth]{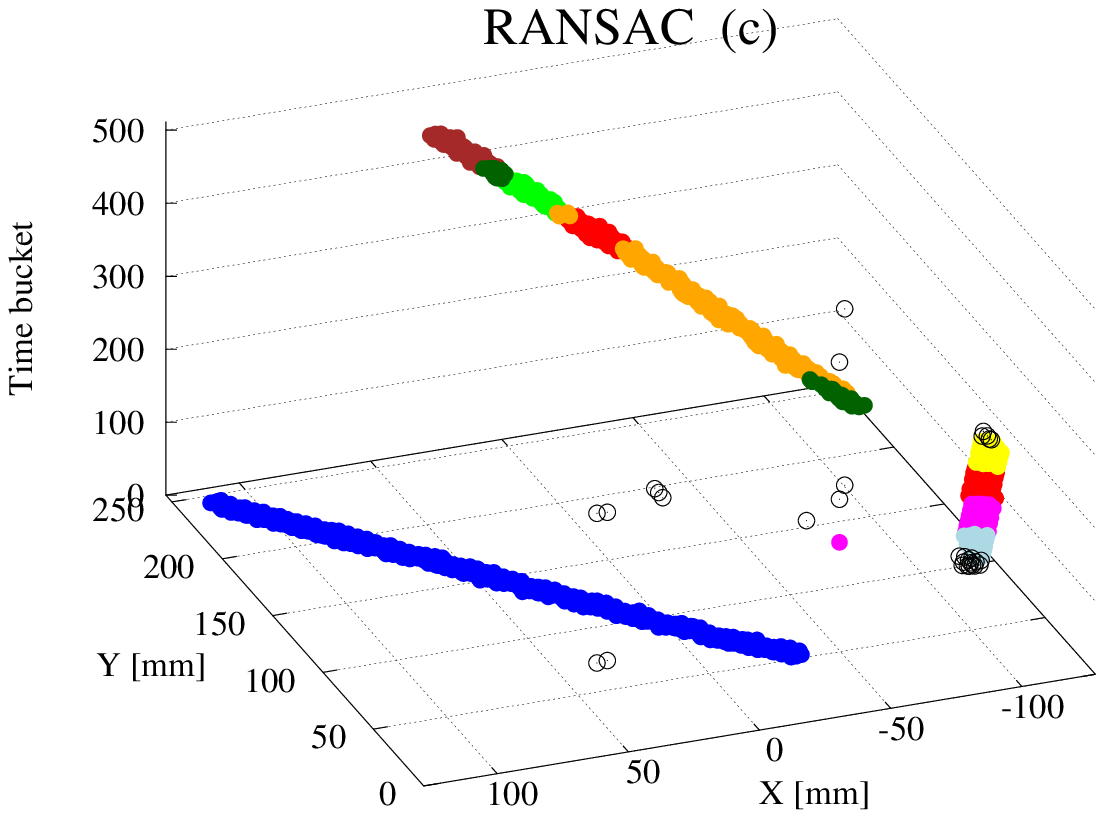}
\includegraphics[width=0.45\textwidth]{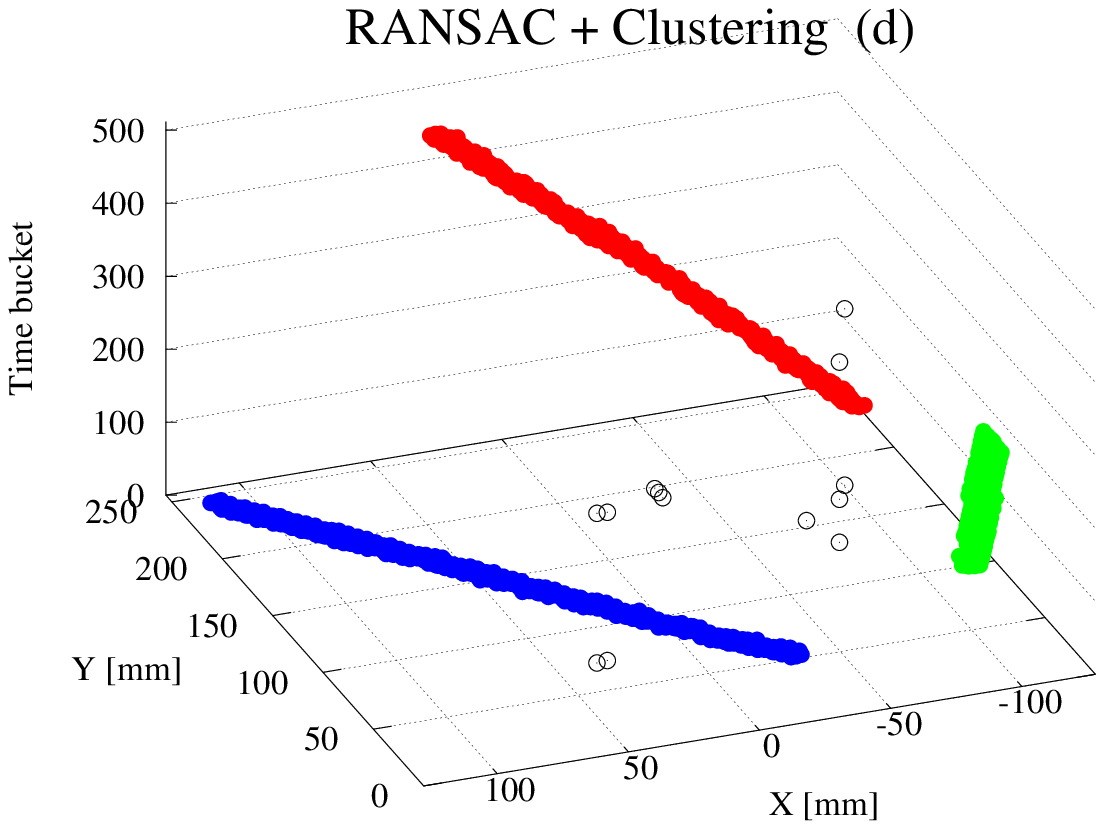}
\caption{\label{multiple} (color online) Examples of track detection errors using RANSAC and the respective solution using clustering. (a) and (b) are experimental data from TexAT, while  (c) and (d) are from ACTAR. In (a), RANSAC has problems for describing short tracks and connects data points from distinct tracks. In (c), RANSAC fails the track detection and over predict the number of models.  }
\end{figure*}

The hit pattern from a nuclear reaction in a TPC contains multiple instances of the same structure (different tracks). Given that  RANSAC-like algorithms are designed to extract a single model,  the estimators need modifications to deal with both gross outliers (noise) and pseudo-outliers (inliers to a different structure). An alternative is to use  the robust estimator sequentially and remove the inliers from the point cloud as each model instance is detected. However, an inaccurate inlier detection for one track contributes heavily to an instability in the fitting of the remaining structures. For instance, when the algorithm selects  hits from different tracks (intersecting model) as inliers, the fitting for all data structures usually fails. Fig.~\ref{multiple}(a) and (c) shows two examples where the  sequential RANSAC provides a wrong estimation for the track fitting. Several methods have been proposed to solve these common problems with RANSAC-like algorithms \cite{1530351,1640882,10.1007/978-3-540-88682-2_41}. Most of these methods rely in a modified random sampling processes and data clustering. The sampling routine can be optimized to generate more useful hypotheses by selecting with a higher probability the hits in the nearest point-cloud region. For example, a first point is sampled uniformly, while the others are sampled with a certain probability distribution (e.g., Eq.~(\ref{nusampling})). This method reduce the problems with intersecting models that may wrongly connect points belonging from two (or more) independent tracks. After all the hypotheses are tested in sequential mode, the particle tracks can be segmented in several groups of inliers or clusters which are very similar. A naive solution for this problem is to increase the threshold parameter, but  the efficiency for the detection of short tracks and the reaction vertex is strongly affected (as it was explained above). A better method is to merge the similar models extracted from the robust estimator by performing an agglomerative clustering. The latter procedure is the basic principle of the J-Linkage algorithm. Fig.~\ref{multiple}(b) shows an example of a successful detection of short tracks with J-Linkage when the sequential RANSAC fails the particle tracking and the vertex reconstruction. 

Similarly, the other routines here presented can be improved by including a clustering process after the model estimation. The first step consist in  classifying the clusters with respect the degree of overlap. The cohesion or separation between clusters can be estimated with the silhouette index \cite{ROUSSEEUW198753}
\begin{equation}
 s(i)=\frac{b(i)-a(i)}{\max \left \{ a(i),b(i) \right \}},
\end{equation}
where $a(i)$ is the mean distance between the point $i$ and all other data points in the same cluster, and $b(i)$ is  the smallest mean distance of the point $i$ to all points in any other cluster. The silhouette index ranges from $-1$ (incorrect clustering) to $+1$ (highly dense clustering).  $s(i)$  values around zero are an indication of overlapping clusters. The relative angle between fitted models provide an extra evaluation for the clusters cohesion. Thus, two clusters belong to the same track if the silhouette index is around zero and their relative angle is small. Figs.~\ref{multiple}(c) and (d) show a comparison of RANSAC with/without using clustering for the tracking with  data from  ACTAR. Sequential RANSAC with uniform sampling generate several incorrect models to describe the data and clearly fails the track fitting. The clustering allows to identify the intersecting models and to agglomerate the clusters that belong to the same track.

\subsection{Efficiency of the algorithms}
\begin{figure}[!ht]
\centering
\includegraphics[width=0.65\textwidth]{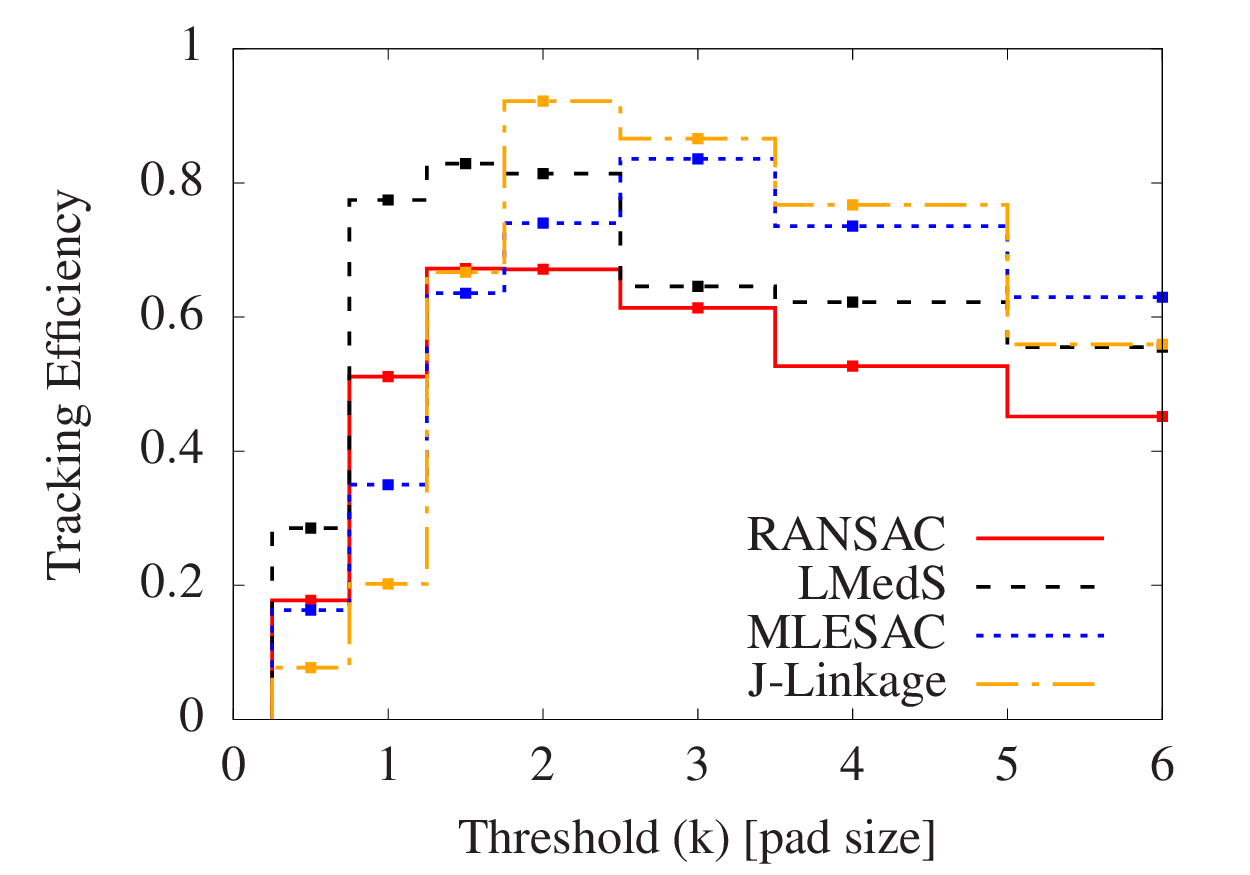}
\caption{\label{effi} (color online) Tracking efficiency as a function of the distance threshold. A threshold parameter of about 2 times the pad size provides good results in our tracking algorithms. The systematic uncertainty is about 10\% in all cases.  }
\end{figure}

In this work, the tracking efficiency was defined as the ratio of the number of good fits and the total number of detected tracks. This parameter was investigated with scattering data from the AT-TPC. Given that the beam was a H$_2$ molecule, most of the events have 3 or more tracks with the same vertex. This makes the present data an excellent test to evaluate the track fitting and vertex reconstruction. In order to test the performance of each algorithm, no clustering was included after the model estimation. As the distance threshold is the most sensitive parameter for the RANSAC-like codes, the tracking efficiency  was obtained as a function of this parameter, as it is shown in Fig.~\ref{effi}. The best efficiency achieved for each algorithm is presented in Table~\ref{table2} .

\begin{table}[!ht]
\caption{\label{table2} Best efficiency ($\varepsilon$) achieved for each algorithm. The systematic uncertainty is about 10\% in all cases. The distance threshold  is given in units of the pad size.   }
\def\arraystretch{1.0}%  1 is the default, change whatever you need
\centering
\begin{tabular}{ccc}
\hline \hline 
Algorithm & $\varepsilon$ [\%] &Threshold [pad size]
      \\ \hline \hline
RANSAC & 67 & 1.5 \\
LMedS & 82 & 1.5 \\
MLESAC & 83 & 3.0 \\
J-Linkage & 92 & 2.0 \\
    \hline \hline    
\end{tabular}
\end{table}

A maximum efficiency of 67\% was obtained with RANSAC for thresholds between 1 and 2 times the pad size. For larger  distance thresholds, the efficiency decrease since the short tracks from $\alpha$ recoiling particles are not detected. LMedS achieve a better efficiency (82\%) with a threshold of 1.5 times the pad size. The smallest median value has a better performance for detecting inliers than the sharp cost function used by RANSAC. MLESAC requires a larger  distance threshold to achieve the maximum efficiency, 83\% at 3 times the pad size. A possible reason for this large threshold is due to the $\sigma$ value ($T=1.96\sigma$) assumed for the inlier probability distribution (see Eq.~\ref{mlesacp}). This also suggest that the parameter $\sigma$ can be adjusted to improve the performance of MLESAC. The best tracking efficiency was obtained with J-Linkage, 92\% for a threshold of 2 times the pad size. This good result is mainly because of the non-uniform sampling used in the algorithm. 

\begin{figure}[!ht]
\centering
\includegraphics[width=0.65\textwidth]{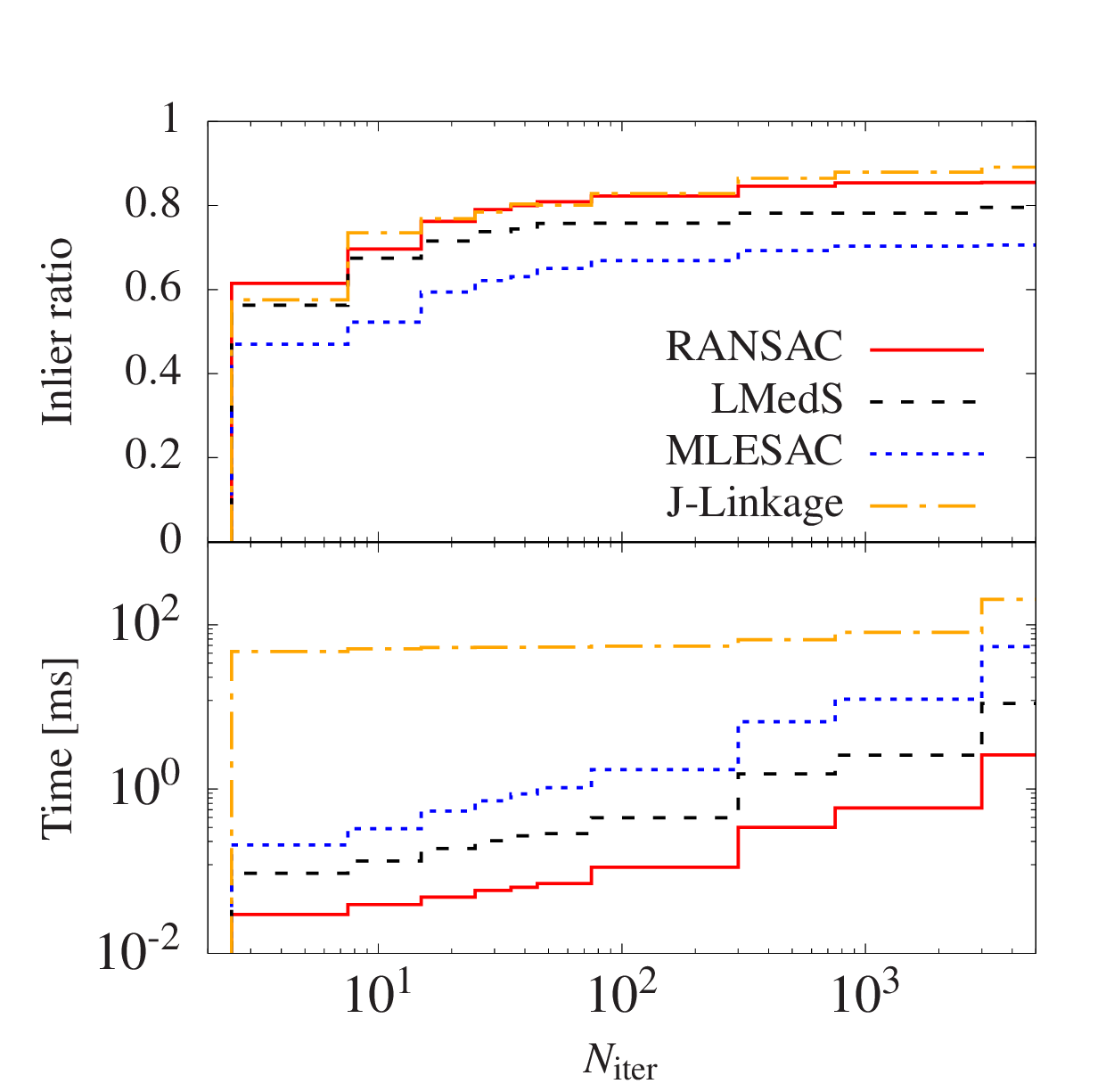}
\caption{\label{runtime} (color online) Inlier ratio (top) and computing time per event (bottom) as a function of the number of iterations in the algorithms. }
\end{figure}

The inlier ratio, $n_\text{inlier}/N_\text{tot}$, was obtained as a function of the number of iterations ($N_\text{iter}$), see Fig.~\ref{runtime}. The four algorithms already converge after 100 iterations. RANSAC and J-Linkage extract the largest number of inliers, while MLESAC has a more strict outlier rejection condition. The runtime per event was also extracted as a function of number of iterations. All the tests were performed on an AMD Ryzen 5 2400G CPU @ 3.9~GHz processor. The computing time for RANSAC is below 1~ms even for a large number of iterations. The robust selection/rejection of data points made by LMedS and MLESAC, increase the runtime by a factor of 3 and 10, respectively. The runtime of J-Linkage is about 0.1~s and it is almost no sensitive to the number of iterations. In this case, the computing time is dominated by the kd-tree routine which is applied for the nearest neighbor and range search at the beginning of each event.

\section{Conclusions}
A tracking algorithm based on  consensus robust estimators was implemented for the analysis of TPC experiments. The routines RANSAC, MLESAC, LMedS and J-Linkage were successfully tested in this work. These type of algorithms provide a simple and efficient method for particle tracking  and reconstruction of the kinematics of nuclear reactions in 3D. The robust estimators beyond RANSAC here implemented make the tracking algorithm less sensitive to noise (and pseudo outliers) and as well as to the parameters of the code. \par
A non-uniform random sampling to select the neighboring points with higher probability  was implemented to reduce the number of intersecting models. Very good results were obtained in all the tests with experimental data, in particular for events containing short tracks. A new weighted sampling method that includes the charge of each hit will be combined with the present sampling routine to improve the generated hypotheses  and reduce the number of iterations. \par
A clustering process after the robust estimation allows to identify similar models that belong to the same track or reject the models that have a wrong prediction.  Excellent results were obtained in the test for track fitting, especially for thick and very dense tracks. However, the clustering step increase considerably the computing time for each event and it should be evaluated whether it is required or not for each application.\par
The test data sets used in this work are experimental data from the AT-TPC, ACTAR and TexAT. This shows the versatility of our tracking algorithm to analyze data taken with any AT detector. Further developments for a detailed particle tracking in complex systems (e.g., including a magnetic field) will be required for application in future projects.

\section*{Acknowledgments}
This work was financially  supported by Fundaç\~ao de Amparo a Pesquisa do Estado de S\~ao Paulo (FAPESP) under Grant Nos.~2018/04965-4, 2016/17612-7  and 2019/07767-1. G.F.F. thanks to Comissão Nacional de Energia Nuclear (CNEN) for the financial support within the MSc. scholarship program. We thank Y.~Ayyad (NSCL) for providing us the experimental data from the AT-TPC, also to B.~Mauss (RIKEN) for the data taken with ACTAR and S.~Ahn (TAMU) for the TexAT data.

%%%%%====================================================================================
\bibliography{bibliography}  % input bibliography
%%%%%====================================================================================

\end{document}